\documentclass[A4,12pt,epsf]{article}

\usepackage{multirow}
\usepackage{amssymb}

\newcommand{\be}{\begin{equation}}
\newcommand{\ee}{\end{equation}}
\newcommand{\bea}{\begin{eqnarray}}
\newcommand{\eea}{\end{eqnarray}}

\begin{document}
\begin{titlepage}
\begin{flushright}
  KIAS-P05020\\
  KUNS-1957
\end{flushright}

\begin{center}
\vspace*{10mm}

{\LARGE \bf Lepton masses and mixing angles from 
heterotic orbifold models}
\vspace{12mm}

{\large
Pyungwon~Ko\footnote{E-mail address: pko@muon.kaist.ac.kr},
Tatsuo~Kobayashi\footnote{E-mail address:
  kobayash@gauge.scphys.kyoto-u.ac.jp}
~and~~Jae-hyeon~Park\footnote{E-mail address:
jhpark@kias.re.kr}
}
\vspace{6mm}


{\it $^{1,3}$School of Physics, KIAS, 
Cheongnyangni-dong, Seoul, 130--722 Korea}\\[1mm]

{\it $^2$Department of Physics, Kyoto University,
Kyoto 606-8502, Japan}\\[1mm]


\vspace*{15mm}

\begin{abstract}
We systematically study the possibility for
realizing  realistic values of lepton mass ratios and 
mixing angles 
by using only renormalizable Yukawa couplings derived from the
heterotic $Z_6$-I orbifold.
We assume one pair of up and down sector Higgs fields.
We consider both the Dirac neutrino mass scenario and 
the seesaw scenario with degenerate right-handed 
majorana neutrino masses.
It is found that realistic values of 
the charged lepton mass ratios, $m_e/m_\tau$ and 
$m_\mu/m_\tau$, the neutrino mass squared difference ratio, 
$\Delta m^2_{31}/\Delta m^2_{21}$, and the lepton
mixing angles can be obtained in certain cases.

\end{abstract}

\end{center}
\end{titlepage}

\section{Introduction}

Understanding the origin of fermion masses and mixing 
angles is one of the most important issues in particle physics.
Quark masses and mixing angles, and charged lepton masses
are well known.
They have hierarchical structures.
Recently, neutrino oscillation experiments have provided
us with information about neutrino mass squared differences and 
mixing angles \cite{Maltoni:2004ei}.
Two of these mixing angles are large in contrast to the small 
quark mixing angles.
Within the framework of the standard model and its 
extensions, a fermion acquires its mass from the Yukawa coupling
with the electroweak Higgs fields.
For right-handed neutrinos, majorana masses are 
additional sources of masses and mixing angles.
Thus, it is important to study Yukawa couplings and 
right-handed majorana neutrino masses derived from 
an underlying theory.

Superstring theory is a promising candidate for a
unified theory including gravity.
For consistency, superstring theory requires 10D
space-time.
That is, it predicts 6D extra space other than 
our 4D space-time, and 
such a space must be compact.
This 6D compact space is very important from 
the viewpoint of string phenomenology.
In general, it is an origin of fermion flavor structure.
That is, the flavor structure derived from a string model
depends on the geometrical aspects of the 6D compact space, 
and in principle one can calculate Yukawa couplings 
for a given compact space with known geometry.
Geometrical aspects of the 6D compact space also provide 
us with selection rules for allowed Yukawa couplings.
In several cases, such selection rules are so severe
that off-diagonal Yukawa couplings are not allowed for 
one Higgs field, because different families are 
discriminated by quantum numbers, which originate from 
geometrical aspects of the 6D compact space.

Among several types of string models, 
heterotic orbifold models \cite{Dixon:jw} 
and intersecting D-brane models are particularly interesting,\footnote{
See for a review of intersecting D-brane models, 
e.g. Ref.~\cite{Blumenhagen:2005mu} and 
references therein.}
to realize realistic Yukawa couplings.
They have localized modes in the 6D compact space, 
and their Yukawa couplings can be 
calculated \cite{Dixon:1986qv,Hamidi:1986vh,
Burwick:1990tu,Kobayashi:2003vi}, 
because their 6D geometry is not complicated.\footnote{
See for calculations of Yukawa couplings in intersecting 
D-brane models Refs.~\cite{Cremades:2003qj,Cvetic:2003ch,
Abel:2003vv}.}
Indeed, a Yukawa coupling among localized modes 
has a suppression factor depending on their 
distances \cite{Dixon:1986qv,Hamidi:1986vh,Ibanez:1986ka,
Burwick:1990tu,Kobayashi:2003vi}.
That can explain suppressed Yukawa couplings.
Therefore, it is very important to study 
possibilities for deriving realistic fermion masses 
and mixing angles from heterotic orbifold models 
and intersecting D-brane models.
The number of 6D $Z_N$ orbifolds \cite{Dixon:jw,Ibanez:1987pj,
Katsuki:1989bf} and $Z_N \times Z_M$ orbifolds \cite{Font:1988mk}, 
especially with 
4D $N=1$ supersymmetry, is finite, i.e.
$Z_3,Z_4,Z_6$-I, $Z_6$-II, $Z_7$, $Z_8$-I, 
$Z_8$-II, $Z_{12}$-I, $Z_{12}$-II, and 
$Z_N \times Z_M$ for $(N,M)=(2,2),(2,3),(2,4),(2,6),
(3,3), (3,6), (4,4), (6,6)$.
(See Refs.~\cite{Ibanez:1987sn} for examples of explicit models,
and Refs.~\cite{Kobayashi:2004ud,Forste:2004ie,
Kobayashi:2004ya,Buchmuller:2004hv} for recent model buildings.)
In particular, non-prime order orbifolds would be 
interesting, because they allow off-diagonal 
couplings \cite{Kobayashi:1991rp,Casas:1991ac}.
On each orbifold, all the fixed points are known, where 
light modes can be localized.
Thus, a systematical analysis on Yukawa matrices, 
which can be derived from heterotic 
orbifold models, would be possible, while
the number of intersecting D-brane configurations 
seems to be infinity.
Indeed, such studies have been done for the quark 
sector \cite{Casas:1992zt,Ko:2004ic}.
In particular, in Ref.~\cite{Ko:2004ic} 
possibilities for leading to the realistic mixing angle 
$V_{cb}$ and mass ratios $m_c/m_t$ and $m_s/m_b$ from 
the $Z_6$-I orbifold have been shown in the case with 
the minimal number of Higgs fields.

However, such a study considering mixing angles has not been done for 
the lepton sector.
In this paper, we study lepton mass ratios and 
mixing angles derived from the $Z_6$-I orbifold.
Commonly, the smallness of neutrino masses is
explained in two ways.
One is the Dirac neutrino mass scenario, 
that is, Yukawa couplings between neutrino and 
Higgs fields are strongly suppressed for some reason.
Since a Yukawa coupling is suppressed 
depending on the size of extra dimensional space 
in an heterotic orbifold model, strongly suppressed 
Yukawa couplings can be obtained in the case that 
an extra dimensional space is large compared with 
the string scale.
The other is the seesaw scenario \cite{seesaw}, which requires 
right-handed majorana neutrino masses at the 
intermediate scale between the weak scale and 
the Planck scale.\footnote{In Ref.~\cite{Giedt:2005vx}, 
realization of the minimal seesaw mechanism has been examined 
in explicit $Z_3$ orbifold models, and its difficulty has been shown.} 
Within the framework of string theory, 
the natural mass scale is just the string scale, 
and such an intermediate mass scale should be obtained 
through vacuum expectation values (VEVs) of some scalar fields.
However, that is quite model-dependent.
Here we study both the Dirac scenario and 
the seesaw scenario, and for the latter case we assume  
the right-handed majorana neutrino mass matrix 
to be proportional to the identity matrix with 
a universal mass scale.
With this, we consider all of the possible assignments 
of leptons to fixed points, examine Yukawa 
terms allowed by the selection rule and calculate 
their Yukawa matrices varying orbifold moduli parameters.
In practice, the number of 
relevant moduli parameters is two in our models.
Then we try to fit the Yukawa matrices by these two moduli 
parameters in order to get realistic values of 
six observables, that is,
lepton mass ratios and mixing angles.

This paper is organized as follows.
In section 2, we give a brief review on 
fixed points on the $Z_6$-I orbifold and 
the corresponding twisted states.
Also, their selection rules for allowed Yukawa couplings and 
the strength of Yukawa couplings are 
reviewed.
In section 3, we study systematically 
the possibility for realizing realistic lepton masses and mixing 
angles by using only renormalizable Yukawa  couplings 
derived from  $Z_6$-I orbifold models.
We assume one pair of up and down sector Higgs fields.
We consider the Dirac neutrino mass scenario in section 3.1.
In section 3.2, we perform the same analysis for 
the seesaw scenario.
Section 4 is devoted to conclusion and discussions.

\section{Orbifold models and selection rule}

Here we give a brief review on heterotic orbifold models, 
that is, the structure of fixed points on orbifolds, 
the selection rule for allowed Yukawa couplings and 
their Yukawa coupling strength.
In particular, we concentrate our attention on
$Z_6$-I orbifold models.
(See for details of generic $Z_N$ orbifolds 
Refs.~\cite{Kobayashi:1991rp,Casas:1991ac}.)

\subsection{Fixed points and twisted sectors}

An orbifold is defined as a division of a torus
by a discrete rotation, i.e., a twist $\theta$.
The 6D $Z_6$-I orbifold is obtained by dividing 
$T^6$ by the twist $\theta$, whose eigenvalues are 
$\mathrm{diag} (e^{2\pi i /6},  e^{2\pi i /6},  e^{2\pi i /3} )$, 
that is, a direct product of two 2D $Z_6$ orbifolds 
and a 2D $Z_3$ orbifold.
The 2D $Z_6$ orbifold is obtained e.g. through  
dividing $R^2$ by the $G_2$ lattice and its automorphism, 
that is, the Coxeter element of $G_2$ algebra, 
which transforms the $G_2$ simple roots $e_1$ and $e_2$, 
\begin{equation}
\theta e_1 \rightarrow -e_1-e_2,  \qquad \theta e_2 \rightarrow 3e_1 +2e_2,
\end{equation}
i.e., the $Z_6$ twist.
Similarly, we can obtain the 2D $Z_3$ orbifold by 
dividing $R^2$ by the $SU(3)$ root lattice and its 
Coxeter element, 
which transforms the $SU(3)$ simple roots
$e_5$ and $e_6$,
\begin{equation}
\theta e_5 \rightarrow e_6,  \qquad \theta e_6 \rightarrow -e_5 -e_6,
\end{equation}
that is, the $Z_3$ rotation.

There are two types of closed strings on an orbifold.
One is the untwisted string, which is already closed on a torus
before orbifold twisting.
The other is the twisted string, which is a localized mode
we are interested in.
A twisted string has the following boundary condition, 
\begin{equation}
X^i(\sigma =  2 \pi) = (\theta^k X)^i(\sigma = 0) + v^i,
\end{equation}
where the shift vector $v^i$ is on the torus lattice $\Lambda$.
That is, the center of mass of a $\theta^k$ twisted 
string is localized at a fixed point $f$, which is
defined as 
\begin{equation}
f^i = (\theta^k f)^i + v^i.
\end{equation}
The fixed point $f$ is represented by the corresponding
space group element $(\theta^k,v^i)$.
The fixed points, which differ by lattice vectors, 
correspond to equivalent fixed points.
That implies that $(\theta^k,v^i)$ is equivalent to 
$(\theta^k,v^i +(1 -\theta^k)\Lambda)$.
Hereafter the $\theta^k$-twisted sector is denoted by 
$\hat T_k$.

The 2D $Z_3$ orbifold has the following 
three fixed points under $\theta$, 
\begin{equation}
g^{(0)}_{Z_3,1}=(0,0), \qquad
g^{(1)}_{Z_3,1}=(2/3,1/3),  \qquad
g^{(2)}_{Z_3,1}(1/3,2/3),
\end{equation}
in the $SU(3)$ simple root basis, and 
each of these is represented by the space group element as
\begin{equation}
g^{(n)}_{Z_3,1}:~~(\theta,n e^1),
\end{equation}
where $n=0,1,2$, up to $(1 - \theta)\Lambda_{SU(3)}$.
The corresponding twisted ground states are denoted by
$|g^{(n)}_{Z_3,1} \rangle$ with $n=0,1,2$.

Similarly, we can obtain fixed points on the 2D 
$Z_6$ orbifold.
The $\theta$ twisted sector on the 2D $Z_6$ 
orbifold has only one fixed point, 
\begin{equation}
g^{(0)}_{Z_6,1} = (0,0),
\end{equation}
in the $G_2$ simple root basis.
The $\theta^2$ twisted sector  has three fixed points, 
\begin{equation}
g^{(0)}_{Z_6,2} = (0,0), \qquad g^{(1)}_{Z_6,2} = (0,1/3), \qquad
g^{(2)}_{Z_6,2} = (0,2/3) .
\end{equation}
Note that these fixed points are defined up to the $G_2$ lattice.
For example, one can use the fixed point 
$g^{(2)}_{Z_6,2} = (1,2/3)$, which is equivalent to $(0,2/3)$.
The corresponding three twisted ground states are denoted by
$| g_{Z_6,2}^{(i)} \rangle$.
However, not all of the three points $g_{Z_6,2}^{(i)}$ are
fixed points of $\theta$.
While $g^{(0)}_{Z_6,2}$ is also a fixed point of 
the twist $\theta$,
the other two fixed points $g^{(1)}_{Z_6,2}$ and $g^{(2)}_{Z_6,2}$ are
transformed to each other by $\theta$.
Since physical states are constructed as $\theta$-eigenstates, 
we take linear combinations of states corresponding to
$g^{(1)}_{Z_6,2}$ and $g^{(2)}_{Z_6,2}$ as
\cite{Kobayashi:1990mc,Kobayashi:1991rp}
\begin{equation}
| g^{(1)}_{Z_6,2}; \pm 1 \rangle \equiv \frac{1}{\sqrt{2}}
\left(| g^{(1)}_{Z_6,2} \rangle
\pm | g^{(2)}_{Z_6,2} \rangle \right),
\label{lin-comb-2}
\end{equation}
with the eigenvalues $\gamma = \pm 1$, while
the state $| g^{(0)}_{Z_6,2} \rangle $ corresponding to
the fixed point $g^{(0)}_{Z_6,2}$ is by itself a $\theta$-eigenstate.

The $\theta^3$ twisted sector has four fixed points,
\begin{eqnarray}
& & g^{(0)}_{Z_6,3} = (0,0), \qquad g^{(1)}_{Z_6,3} = (0,1/2),
\nonumber \\
& & g^{(2)}_{Z_6,3} = (1/2,0), \qquad  g^{(3)}_{Z_6,3} = (1/2,1/2) .
\end{eqnarray}
Recall that these fixed points are defined up to the $G_2$ lattice.
For instance, the fixed point $g^{(1)}_{Z_6,3} = (1,1/2)$ is 
equivalent to $(0,1/2)$.
Not all of the four points are fixed points of the twist $\theta$.
The $\theta$-eigenstates for each $G_2$ part are obtained as
\begin{equation}
 | g^{(0)}_{Z_6,3} \rangle, \qquad
 | g^{(1)}_{Z_6,3}; \gamma \rangle \equiv  \frac{1}{\sqrt{3}}
\left(| g^{(1)}_{Z_6,3} \rangle
+ \gamma  | g^{(2)}_{Z_6,3} \rangle + \gamma^2  | g^{(3)}_{Z_6,3}
 \rangle \right),
\end{equation}
where $\gamma = 1, \omega, \omega^2$ with $\omega = e^{2 \pi i/3}$.

A fixed point on the 6D
$Z_6$-I orbifold is obtained as a direct product of
three fixed points,
each coming from one of the two 2D $Z_6$ orbifolds and
the 2D $Z_3$ orbifold.
The corresponding twisted ground state is
obtained in the same manner.
The $\theta$ twisted sector has the following ground states,
\begin{equation}
|g^{(0)}_{Z_6,1}\rangle \otimes |g^{(0)}_{Z_6,1} \rangle
\otimes | g^{(i)}_{Z_3,1} \rangle ,
\end{equation}
for $i=0,1,2$.
The $\theta^2$ twisted sector has the following
ground states,
\begin{eqnarray}
& & | g^{(0)}_{Z_6,2}\rangle \otimes | g^{(0)}_{Z_6,2} \rangle
\otimes | g^{(j)}_{Z_3,2} \rangle,  \nonumber \\
& & | g^{(1)}_{Z_6,2};\gamma \rangle \otimes | g^{(0)}_{Z_6,2} \rangle
\otimes | g^{(j)}_{Z_3,2} \rangle,  \nonumber \\
& & | g^{(0)}_{Z_6,2}\rangle \otimes | g^{(1)}_{Z_6,2}; \gamma' \rangle
\otimes | g^{(j)}_{Z_3,2} \rangle,  \\
& & | g^{(1)}_{Z_6,2};\gamma\rangle \otimes | g^{(1)}_{Z_6,2}; \gamma' \rangle
\otimes | g^{(j)}_{Z_3,2} \rangle,  \nonumber
\end{eqnarray}
for $\gamma, \gamma' = \pm 1$ and $j=0,1,2$.
The $\theta^3$ twisted sector has the following 
ground states,
\begin{eqnarray}
& & | g^{(0)}_{Z_6,3}\rangle \otimes | g^{(0)}_{Z_6,3}\rangle,
\nonumber \\
& & | g^{(1)}_{Z_6,3};\gamma\rangle \otimes | g^{(0)}_{Z_6,3}\rangle,
\nonumber \\
& & | g^{(0)}_{Z_6,3}\rangle \otimes | g^{(1)}_{Z_6,3}; \gamma' \rangle,  \\
& & | g^{(1)}_{Z_6,3};\gamma\rangle \otimes | g^{(1)}_{Z_6,3};
\gamma' \rangle, \nonumber
\end{eqnarray}
where $\gamma, \gamma' = 1, \omega, \omega^2$.

\subsection{Selection rule}

Here we give a brief review on the selection rule for Yukawa couplings
in orbifold models.
(See for their details Refs.~\cite{Kobayashi:1991rp,Casas:1991ac}.)
The fixed point $f$ of the $\theta^k$ twisted sector is 
denoted by its space group element,
$(\theta^k, (1-\theta^k) f)$, as said in the
previous subsection.
Thus, the three states corresponding to the three fixed points
$(\theta^{k_i},(1-\theta^{k_i})f_i)$ for $i=1,2,3$ can couple if the
product of their space group elements
$\prod_i (\theta^{k_i},(1-\theta^{k_i})f_i)$  is
equivalent to identity.
That implies the space group selection rule for allowed Yukawa couplings 
requires \cite{Dixon:1986qv}
\begin{equation}
\prod_i \left( \theta^{k_i},(1-\theta^{k_i})(f_i+\Lambda) \right) =
(1,0),
\end{equation}
because the fixed point $(\theta^k, (1-\theta^k) f)$
is equivalent to $(\theta^k, (1-\theta^k) (f+ \Lambda) )$.
This space group selection rule includes the point group 
selection rule, that is, the product of twists must be 
identity, $\prod_i \theta^{k_i} = 1$.
In $Z_6$-I orbifold models, 
the point group selection rule and $H$-momentum
conservation \cite{Friedan:1985ge}
allow only the following couplings \cite{Kobayashi:1990mc},
\begin{equation}
\hat T_1 \hat T_2 \hat T_3, \qquad \hat T_2 \hat T_2 \hat T_2 .
\end{equation}

The space group selection rule for the 2D $Z_3$ orbifold is 
simple.
For a $\hat T_2 \hat T_2 \hat T_2$ coupling, 
three states
corresponding to fixed points $g^{(i_1)}_{Z_3,2}$,
$g^{(i_2)}_{Z_3,2}$ and $g^{(i_3)}_{Z_3,2}$, can couple when
the following equation is satisfied,
\begin{equation}
i_1 + i_2 + i_3 = 0 ~~~({\rm mod~~~}3).
\label{z3-symm}
\end{equation}
Also, for a $\hat T_1 \hat T_2 \hat T_3$ coupling, 
three states can couple when the fixed point of $\hat T_1$ is 
the same as that of $\hat T_2$.
Thus, the space group selection rule of 
the 2D $Z_3$ part allows only diagonal couplings, 
that is, this part is not relevant to our purpose of deriving
realistic Yukawa matrices with  
non-vanishing mixing angles in the case with the 
minimal number of Higgs fields.
For a while, we will assume that all relevant states correspond to 
the same fixed point on the 2D $Z_3$ orbifold.
There is another possibility for a $\hat T_2 \hat T_2 \hat T_2$
coupling, that is, the three states correspond to 
different fixed points on the 2D $Z_3$ orbifold, 
and in this case, we have a suppressed Yukawa coupling
depending on the volume of the 2D $Z_3$ orbifold.
We will give a comment on it later.

We can discuss the space group selection rule for the 
2D $Z_6$ part.
For $\hat T_2 \hat T_2 \hat T_2$ couplings, 
the space group selection rule on the 2D $Z_6$ orbifold 
is exactly the same as that on the 2D $Z_3$ orbifold, 
i.e. eq.(\ref{z3-symm}),  
when we consider the basis of twisted states 
corresponding directly to fixed points.
However, in $Z_6$-I orbifold models, we take linear combinations 
as in Eq.~(\ref{lin-comb-2}).
The space group selection rule for $\hat T_1 \hat T_2 \hat T_3$ 
couplings is non-trivial.
All couplings on the 2D $Z_6$ orbifold are allowed by 
the space group selection rule, because
$(1-\theta)\Lambda = \Lambda$.
Thus, off-diagonal couplings are allowed 
for $\hat T_1 \hat T_2 \hat T_3$ couplings in the 
2D $Z_6$ orbifold part.
Furthermore, the selection rule requires that 
the product of $\theta$-eigenvalues of
the coupling states must be equal to identity, that is,
$\prod \gamma = 1$.
Therefore, the twisted states, which are relevant to 
our purpose, are the single $\hat T_1$ state,
\begin{equation}
| g^{(0)}_{Z_6,1} \rangle \otimes | g^{(0)}_{Z_6,1} \rangle ,
\end{equation}
and the five $\hat T_2$ states,
\begin{eqnarray}
\hat T_2^{(1)} &\equiv&
| g^{(0)}_{Z_6,2} \rangle \otimes | g^{(0)}_{Z_6,2} \rangle ,\nonumber \\
\hat T_2^{(2)} &\equiv&
| g^{(0)}_{Z_6,2} \rangle \otimes | g^{(1)}_{Z_6,2}; +1 \rangle ,
\nonumber \\
\hat T_2^{(3)} &\equiv&
| g^{(1)}_{Z_6,2}; +1 \rangle \otimes | g^{(0)}_{Z_6,2} \rangle ,
\label{T2-states} \\
\hat T_2^{(4,\gamma)} &\equiv&
| g^{(1)}_{Z_6,2}; \gamma \rangle \otimes | g^{(1)}_{Z_6,2};
\gamma^{-1} \rangle , \nonumber
\end{eqnarray}
where $\gamma = \pm 1$, and the six $\hat T_3$ states,
\begin{eqnarray}
\hat T_3^{(1)} &\equiv&
| g^{(0)}_{Z_6,3} \rangle \otimes | g^{(0)}_{Z_6,3} \rangle ,
\nonumber \\
\hat T_3^{(2)} &\equiv&
| g^{(0)}_{Z_6,3} \rangle \otimes | g^{(1)}_{Z_6,3}; +1 \rangle ,
\nonumber \\
\hat T_3^{(3)} &\equiv&
| g^{(1)}_{Z_6,3}; +1 \rangle \otimes | g^{(0)}_{Z_6,3} \rangle ,
\label{T3-states} \\
\hat T_3^{(4,\gamma)} &\equiv&
| g^{(1)}_{Z_6,3}; \gamma \rangle \otimes | g^{(1)}_{Z_6,3};
\gamma^{-1} \rangle , \nonumber
\end{eqnarray}
where $\gamma = 1, \omega, \omega^2$.
We have omitted the index for the fixed point on the 
2D $Z_3$ orbifold, because 
we assume all states sit on the same fixed point 
on the 2D $Z_3$ orbifold for the moment.

\subsection{Yukawa couplings}

The strength of a Yukawa coupling has been calculated
by use of 2D conformal field theory.
It depends on distances between fixed points.
The Yukawa coupling strength of the
$\hat T_1 \hat T_2 \hat T_3$ coupling in $Z_6$-I
orbifold models is obtained for the $G_2 \times G_2 $ part 
as \cite{Dixon:1986qv,Hamidi:1986vh,Burwick:1990tu,Casas:1991ac}
\begin{equation}
Y = \sum_{f_{23}=f_2-f_3 +\Lambda} \exp [-\frac{\sqrt 3}{4 \pi} f_{23}^T
M f_{23}],
\label{Yukawa-123}
\end{equation}
up to an overall normalization factor, where
\begin{equation}
M = \left(
\begin{array}{cccc}
R_1^2 & -\frac{3}{2}R_1^2 & 0 & 0 \\ -\frac{3}{2}R_1^2 &
3R_1^2 & 0& 0 \\
0 & 0& R_2^2 & -\frac{3}{2}R_2^2 \\
0 & 0& -\frac{3}{2}R_2^2 & 3R_2^2
\end{array}
\right) ,
\label{M-matrix}
\end{equation}
in the $G_2 \times G_2$ root basis.
Here, $f_2$ and $f_3$ denote fixed points of the
$\hat T_2$ and $\hat T_3$ sectors, respectively, and
$R_i$ corresponds to the radius of the $i$-th torus,
which can be written as the real part of the $i$-th
K\"ahler modulus $T_i$ up to a constant factor.
Here we follow Ref.~\cite{Casas:1991ac} 
for the normalization  of $R_i$.
(See Ref.~\cite{Lebedev:2001qg} for the normalization
of the moduli
such that the transformation
$T_\ell \rightarrow T_\ell +i$ is a symmetry.)
The imaginary parts of $T_i$ also contribute to mass matrices, 
i.e., eigenvalues, mixing angles and CP violating phases.
However, here we consider only the real parts $R_i$ 
for simplicity.
Since the states with fixed points in the same 
conjugacy class contribute to the Yukawa coupling, 
we take summation of these contributions in eq.~(\ref{Yukawa-123}).
However, the states corresponding to the nearest
fixed points $(f_2,f_3)$ contribute dominantly
to the Yukawa coupling for a large value of $R_i$.
Hence, we calculate Yukawa couplings only by 
the contribution due to the nearest
fixed points $(f_2,f_3)$.
Indeed such an approximation is valid when $R_i$  is sufficiently 
large as in the cases we will study in the following sections.

Similarly, the strength of $\hat T_2 \hat T_2 \hat T_2$
Yukawa couplings is obtained 
in the basis of twisted states corresponding 
directly to fixed points as
\begin{equation}
Y = \sum_{f_{23}= f_2-f_3 +\Lambda}
\exp [- \frac{\sqrt 3}{16 \pi}f_{23}^T M
f_{23}],
\label{Yukawa-222}
\end{equation}
where $M$ is the same matrix as Eq.(\ref{M-matrix}).
Here, $f_2$ and $f_3$ denote two of the three fixed points
in the $\hat T_2$ sector.
Recall that when we choose two states, the other state,
which is allowed to couple, is uniquely fixed 
in the basis of states corresponding directly to 
fixed points.

Here, we give a comment on the K\"ahler metric.
For a given $k$, the $\theta^k$ twisted states have the same 
K\"ahler metric, even if they correspond to 
different fixed points.
Thus, the K\"ahler metric is irrelevant to 
mass ratios or mixing angles when 
the three families of leptons with the
same $SU(3)\times SU(2) \times U(1)_Y$ quantum numbers 
are assigned to
states in a single $\hat T_k$ sector.
Indeed this type of assignment is required by 
the point group selection rule in order that 
non-vanishing  mixing angles can be realized.

\section{Lepton masses and mixing angles}

Here we systematically study the possibility for deriving 
realistic lepton masses and mixing angles by use of 
twisted states, their selection rules and the 
Yukawa couplings, which are shown in 
the previous section.
We assume that we obtain the standard gauge group $SU(3) \times SU(2) 
\times U(1)_Y$ and 
three families of leptons in a $Z_6$-I orbifold
model.
Indeed it is quite a nontrivial issue to 
construct a realistic model including the gauge group 
$SU(3) \times SU(2) \times U(1)_Y$ and three families of 
quarks and leptons, but here we just assume them, 
because our purpose is not to construct an explicit model, but 
to show the possibility for leading to realistic mixing angles and 
mass ratios.  
We also assume one pair of up and down
Higgs fields.

Charged lepton masses are well known and shown in 
Table~\ref{tab:input} \cite{Eidelman:2004wy}.
The table also shows the neutrino mass squared differences and 
mixing angles, which are consistent with 
recent experiments on neutrino oscillations \cite{Maltoni:2004ei}.

\begin{table}
  \centering
  \begin{tabular}{cc}
    \hline
    Parameter & Experimental value \\ \hline
    $m_e$ & $(0.51099892 \pm 0.00000004)$ MeV \\
    $m_\mu$ & $(105.658369 \pm 0.000009)$ MeV \\
    $m_\tau$ & $(1776.99 \pm 0.275)$ MeV \\
    $\Delta m^2_{21}$ & $(8.1 \pm 0.3) \times 10^{-5}\ \mathrm{eV}^2$ \\
    $\Delta m^2_{31}$ & $(2.2 \pm 0.3) \times 10^{-3}\ \mathrm{eV}^2$ \\
    $\sin^2 \theta_{12}$ & $0.30 \pm 0.025$ \\
    $\sin^2 \theta_{23}$ & $0.50 \pm 0.065$ \\
    $\sin^2 \theta_{13}$ & $\le 0.014$ \\
    \hline
  \end{tabular}
  \caption{Experimental values of lepton masses and mixing angles.}
  \label{tab:input}
\end{table}

\subsection{Dirac neutrino mass scenario}

First we consider the Dirac neutrino mass scenario.
The relevant terms in the superpotential are 
\begin{equation}
  W \supset H_u L_i (Y_\nu)_{ij} N_j - H_d L_i (Y_e)_{ij} e^c_j ,
\end{equation}
where $H_u$ and $H_d$ are the up- and down-sector Higgs fields,
$L_i$ are the SU(2) doublet leptons, $N_i$ are the gauge singlet neutrinos,
and $e^c_i$ are the SU(2) singlet charged leptons.
After replacing the Higgs fields with their VEVs, $v_u$ and $v_d$, 
leptons gain the mass terms,
\begin{equation}
  W \supset - \nu_i (m_D)_{ij} N_j - e_i (m_e)_{ij} e^c_j ,
\end{equation}
where $(\nu_i, e_i) = L_i$, $(m_D)_{ij} = (v_u Y_\nu)_{ij}$, and 
$(m_e)_{ij} = (v_d Y_e)_{ij}$.
The mass matrices can be diagonalized as
\begin{equation}
  \label{eq:biunitary}
    m_D = U_\nu m^\mathrm{diag}_D V_\nu^\dagger , \quad
    m_e   = U_e m^\mathrm{diag}_e V_e^\dagger .
\end{equation}
In terms of these unitary matrices, the lepton mixing matrix is
defined as
\begin{equation}
  \label{eq:pmns}
  U_\mathrm{MNS} \equiv U_e^T U_\nu^* .
\end{equation}

We consider all possible assignments of 
$L_i, N_i, e^c_i$ and $H_{u,d}$ to the twisted states
shown in the previous section.
The point group selection rule implies that we can obtain 
non-vanishing mixing angles only when each species of
$L_i$, $N_i$, and $e^c_i$, belongs to the same twisted sector.
The $\hat T_1$ sector has a single state for the 
$T^2/Z_6 \times T^2/Z_6$ part.
Hence, we have to assign $L_i$, $N_i$ and $e^c_i$ to 
$\hat T_2$ or $\hat T_3$ to obtain non-vanishing 
mixing angles, while the Higgs fields can be 
assigned to $\hat T_1$.
Therefore, there are five classes of assignments, which are shown in 
Table~\ref{tab:assign}. 
For each of possible assignments, we examine the selection rule 
for allowed Yukawa couplings and calculate Yukawa couplings 
as functions of $R_1$ and $R_2$.
Then, varying these two parameters  $R_1$ and $R_2$, 
we try to fit the charged lepton mass ratios $m_e/m_\tau$ and 
$m_\mu/m_\tau$, and the ratio of neutrino mass squared difference 
$\Delta m^2_{31}/\Delta m^2_{21}$ and mixing angles 
$\theta_{12}$, $\theta_{23}$, 
$\theta_{13}$.
For the ease of presentation, we display
the ratios of experimental values,
which can be immediately obtained from Table~\ref{tab:input}.
\begin{eqnarray}
    ( m_e/m_\tau, m_\mu/m_\tau ) &=& ( 0.000288, 0.0595 ), \\
    \Delta m^2_{31}/\Delta m^2_{21} &=& 27 .
\end{eqnarray}
In particular, we are interested in deriving 
their orders, but not precise values, 
because those values are obtained at the string scale.
However, it is quite non-trivial to fit these six observables, 
$m_e/m_\tau$, $m_\mu/m_\tau$, 
$\Delta m^2_{31}/\Delta m^2_{21}$ 
$\sin^2 \theta_{12}$, $\sin^2 \theta_{23}$, 
$\sin^2 \theta_{13}$, 
only by two parameters, $R_1$ and $R_2$. 
It may be most important to realize the mass ratios of 
charged leptons $m_e/m_\tau$ and  $m_\mu/m_\tau$, 
because their values are measured very precisely by 
experiments.
Hence, the two parameters, $R_1$ and $R_2$, 
are almost fixed in order to fit $m_e/m_\tau$ and  $m_\mu/m_\tau$.
In all the five classes of assignments, we can find cases
that fit the charged lepton mass ratios, but
not in all of them can we fit the neutrino oscillation data
because mass matrix patterns are limited.

\begin{table}[tb]
  \centering
\begin{tabular}{|c|c|c|c|c|c|} \hline
Class & $L$ & $N$ & $e^c$ & $H_u$ & $H_d$ \\ \hline \hline
Assignment 1 & $\hat T_2$ & $\hat T_3$ &  $\hat T_3$ &
 $\hat T_1$ & $\hat T_1$ \\ \hline
Assignment 2 & $\hat T_3$ & $\hat T_2$ &  $\hat T_2$ &
 $\hat T_1$ & $\hat T_1$ \\ \hline
Assignment 3 & $\hat T_2$ & $\hat T_3$ &  $\hat T_2$ &
 $\hat T_1$ & $\hat T_2$ \\ \hline
Assignment 4 & $\hat T_2$ & $\hat T_2$ &  $\hat T_3$ &
 $\hat T_2$ & $\hat T_1$ \\ \hline
Assignment 5 & $\hat T_2$ & $\hat T_2$ &  $\hat T_2$ &
 $\hat T_2$ & $\hat T_2$ \\ \hline
\end{tabular}
  \caption{Five classes of assignments}
  \label{tab:assign}
\end{table}

We show our results in what follows.

\vskip .3cm
{\bf Assignment 1}

In this class of assignments, we can fit only
the mass ratios $m_e/m_\tau$ and  $m_\mu/m_\tau$ 
properly.
For example, we take $(R^2_1,R^2_2)=(26,33)$ in the following 
assignment,
\begin{eqnarray}
(L_1,L_2,L_3) = (\hat T^{(1)}_2,\hat T^{(2)}_2,\hat T^{(4,1)}_2), & & 
(N_1,N_2,N_3) = (\hat T^{(1)}_3,\hat T^{(2)}_3,\hat T^{(3)}_3), 
\nonumber \\
(e^c_1,e^c_2,e^c_3) = (\hat T^{(1)}_3,\hat T^{(2)}_3,\hat T^{(4,1)}_3), & & 
(H_u,H_d) = (\hat T^{}_1,\hat T^{}_1).
\end{eqnarray}
Then, we can realize the experimental values of 
$m_e/m_\tau$ and  $m_\mu/m_\tau$.
However, the neutrino oscillation data cannot be well accommodated.
In this particular assignment, we obtain 
\begin{eqnarray}
\frac{\Delta m^2_{31}}{\Delta m^2_{21}} = 110, & &
    \sin^2 \theta_{12} = 5 \times 10^{-5}, \nonumber \\
\sin^2 \theta_{23} = 6 \times 10^{-3}, & &\sin^2 \theta_{13} = 6 
\times 10^{-11}. 
\end{eqnarray}
Thus, the mass squared difference ratio $\Delta m^2_{31}/\Delta m^2_{21}$
is large, and what is worse is that 
the mixing angles $\theta_{12}$ and $\theta_{23}$ are 
too small.
This is  the character of this class of assignments. 
Namely, we can fit the charged lepton mass ratios 
$m_e/m_\tau$ and  $m_\mu/m_\tau$, but the mixing angles are 
too small except the trivial cases.
This character is shown in Table~\ref{tab:char-D}.
Here and hereafter, by a trivial result we mean that at least 
one of charged leptons is massless, 
charged leptons or neutrinos have degenerate masses, 
or the mixing matrix is trivially the identity matrix. 

\vskip .3cm
{\bf Assignment 2}

In this class of assignments, we can fit only the mass ratios  
$m_e/m_\tau$ and  $m_\mu/m_\tau$ 
properly.
For example, we take $(R^2_1,R^2_2)=(21,26)$ in the following 
assignment,
\begin{eqnarray}
(L_1,L_2,L_3) = (\hat T^{(1)}_3,\hat T^{(2)}_3,\hat T^{(3)}_3), & & 
(N_1,N_2,N_3) = (\hat T^{(1)}_2,\hat T^{(2)}_2,\hat T^{(4,1)}_2), 
\nonumber \\
(e^c_1,e^c_2,e^c_3) = (\hat T^{(1)}_2,\hat T^{(3)}_2,\hat T^{(4,1)}_2), & & 
(H_u,H_d) = (\hat T^{}_1,\hat T^{}_1).
\end{eqnarray}
Then, we can realize the proper mass ratios 
$m_e/m_\tau$ and  $m_\mu/m_\tau$. 
The neutrino masses and mixing angles are not so satisfactory, however.
We obtain 
\begin{eqnarray}
\frac{\Delta m^2_{31}}{\Delta m^2_{21}} = 150, & &
    \sin^2 \theta_{12} = 4 \times 10^{-7}, \nonumber \\
\sin^2 \theta_{23} = 2 \times 10^{-3}, & &\sin^2 \theta_{13} = 9 
\times 10^{-8}. 
\end{eqnarray}
Thus, the mass squared difference ratio $\Delta m^2_{31}/\Delta m^2_{21}$
is large, and 
the mixing angles $\theta_{12}$ and 
$\theta_{23}$ are too small.
That is the character of this class of assignments.
Namely, we can fit the charged lepton mass ratios 
$m_e/m_\tau$ and  $m_\mu/m_\tau$, but the mixing angles are 
too small except the trivial cases.
This character is shown in Table~\ref{tab:char-D}.

\vskip .3cm
{\bf Assignment 3}


In this class of assignments, we can fit only
$m_e/m_\tau$, $m_\mu/m_\tau$, and $\Delta m^2_{31}/\Delta m^2_{21}$ properly.
For example, we take $(R^2_1,R^2_2)=(205,481)$ in the following 
assignment,
\begin{eqnarray}
(L_1,L_2,L_3) = (\hat T^{(2)}_2,\hat T^{(3)}_2,\hat T^{(4,1)}_2), & & 
(N_1,N_2,N_3) = (\hat T^{(1)}_3,\hat T^{(2)}_3,\hat T^{(4,1)}_3), 
\nonumber \\
(e^c_1,e^c_2,e^c_3) = (\hat T^{(1)}_2,\hat T^{(3)}_2,\hat T^{(4,1)}_2), & & 
(H_u,H_d) = (\hat T^{}_1,\hat T^{(2)}_2).
\end{eqnarray}
Then, we can realize the proper mass ratios 
$m_e/m_\tau$ and  $m_\mu/m_\tau$. 
On the other hand, we obtain
\begin{eqnarray}
\frac{\Delta m^2_{31}}{\Delta m^2_{21}} = 19, & &
    \sin^2 \theta_{12} = 3 \times 10^{-4}, \nonumber \\
\sin^2 \theta_{23} = 1.0, & &\sin^2 \theta_{13} = 3
\times 10^{-10}. 
\end{eqnarray}
Thus, the mass squared difference ratio $\Delta m^2_{31}/\Delta
  m^2_{21}$ is compatible with the data, but
the mixing angle $\theta_{12}$ ($\theta_{23}$) is too small (too big).
The sizes of $\sin^2 \theta_{12,23,13}$ vary from case to case, but
they turn out to be either too big $\gtrsim 0.9$ or too small $\lesssim 0.1$
in most cases.
Sometimes they are 0.3 or 0.5, but no case leads to
the right values of the three mixing angles at the same time.
That is the character of this class of assignments.
Namely, we can fit the charged lepton mass ratios 
$m_e/m_\tau$ and  $m_\mu/m_\tau$, and the neutrino mass squared difference
ratio, $\Delta m^2_{31}/\Delta m^2_{21}$, but the mixing angles
are either too big or too small.
This character is shown in Table~\ref{tab:char-D}.

\vskip .3cm
{\bf Assignment 4}

In this class of assignments, we can fit
the neutrino oscillation data as well as
the mass ratios $m_e/m_\tau$ and  $m_\mu/m_\tau$. 
For example, we take $(R^2_1,R^2_2) = (26,21)$ 
in the following assignment,
\begin{eqnarray}
(L_1,L_2,L_3) = (\hat T^{(2)}_2,\hat T^{(3)}_2,\hat T^{(4,-1)}_2), & & 
(N_1,N_2,N_3) = (\hat T^{(2)}_2,\hat T^{(4,1)}_2,\hat T^{(4,-1)}_2), 
\nonumber \\
(e^c_1,e^c_2,e^c_3) = (\hat T^{(1)}_3,\hat T^{(2)}_3,\hat T^{(4,1)}_3), & & 
(H_u,H_d) = (\hat T^{(2)}_2,\hat T^{}_1).
\end{eqnarray}
Then, we can realize the experimental values of 
$m_e/m_\tau$ and  $m_\mu/m_\tau$.
Moreover, in this assignment we obtain 
\begin{eqnarray}
\frac{\Delta m^2_{31}}{\Delta m^2_{21}} = 14, & &
    \sin^2 \theta_{12} = 0.38, \nonumber \\
\sin^2 \theta_{23} = 0.70, & &\sin^2 \theta_{13} = 6.3 
\times 10^{-6}. 
\end{eqnarray}
Thus, this assignment can realize orders of 
six observables only by two parameters.
Almost the same result can be derived from the 
following assignment,
\begin{eqnarray}
(L_1,L_2,L_3) = (\hat T^{(2)}_2,\hat T^{(3)}_2,\hat T^{(4,-1)}_2), & & 
(N_1,N_2,N_3) = (\hat T^{(2)}_2,\hat T^{(4,1)}_2,\hat T^{(4,-1)}_2), 
\nonumber \\
(e^c_1,e^c_2,e^c_3) = (\hat T^{(2)}_3,\hat T^{(3)}_3,\hat T^{(4,1)}_3), & & 
(H_u,H_d) = (\hat T^{(2)}_2,\hat T^{}_1).
\end{eqnarray}
There are also several other assignments leading to similar results with 
smaller values of $\Delta m^2_{31}/\Delta m^2_{21}$
in the range of
\begin{equation}
1< \frac{\Delta m^2_{31}}{\Delta m^2_{21}} < 14.
\end{equation}
In Table~\ref{tab:char-D}, the best fitting result is shown.

\vskip .3cm
{\bf Assignment 5}


In this class of assignments, we can fit only
$m_e/m_\tau$, $m_\mu/m_\tau$, and $\Delta m^2_{31}/\Delta m^2_{21}$ properly.
For example, we take $(R^2_1,R^2_2)=(243,419)$ in the following 
assignment,
\begin{eqnarray}
(L_1,L_2,L_3) = (\hat T^{(1)}_2,\hat T^{(2)}_2,\hat T^{(4)}_2), & & 
(N_1,N_2,N_3) = (\hat T^{(1)}_2,\hat T^{(3)}_2,\hat T^{(4,1)}_2), 
\nonumber \\
(e^c_1,e^c_2,e^c_3) = (\hat T^{(2)}_2,\hat T^{(3)}_2,\hat T^{(4,1)}_2), & & 
(H_u,H_d) = (\hat T^{(2)}_2,\hat T^{(4,1)}_2).
\end{eqnarray}
Then, we can realize the proper mass ratios 
$m_e/m_\tau$ and  $m_\mu/m_\tau$. 
On the other hand, we obtain
\begin{eqnarray}
\frac{\Delta m^2_{31}}{\Delta m^2_{21}} = 28, & &
    \sin^2 \theta_{12} = 1 \times 10^{-3}, \nonumber \\
\sin^2 \theta_{23} = 0.06, & &\sin^2 \theta_{13} = 3
\times 10^{-14}. 
\end{eqnarray}
Thus, the mass squared difference ratio $\Delta m^2_{31}/\Delta
  m^2_{21}$ is compatible with the data, but
the mixing angles $\theta_{12}$ and 
$\theta_{23}$ are too small.
The size of $\sin^2 \theta_{23}$ varies from case to case, but
it turns out to be either too big $\sim 0.9$ or too small $\sim 0.1$
in any case.
That is the character of this class of assignments.
Namely, we can fit the charged lepton mass ratios 
$m_e/m_\tau$ and  $m_\mu/m_\tau$, but $\theta_{12}$ is too small
and $\theta_{23}$ is either too big or too small.
This character is shown in Table~\ref{tab:char-D}.

\vskip .3cm

As a result, we can realize the charged lepton mass ratios, 
the ratio of neutrino mass squared differences and 
the lepton mixing angles in certain cases of Assignment 4, 
but not in the other classes of assignments.
Let us also note that the neutrino
mass spectrum has normal hierarchy
with vanishing or very small lightest neutrino mass
in every case leading to reasonable value of
$\Delta m^2_{31}/\Delta m^2_{21}$.

So far, we have assumed that all states correspond 
to the same fixed point on the 2D $Z_3$ orbifold.
However, that case leads to $(Y_\nu)_{33} =O(1)$ 
which is not realistic.  We need a suppression 
factor of $O(10^{-12}-10^{-13})$
to fit the overall magnitude.
In Assignment 4, the neutrino Yukawa couplings 
originate from $\hat T_2 \hat T_2 \hat T_2$ couplings.
On the 2D $Z_3$ orbifold, Yukawa couplings corresponding to
three different fixed points are allowed, too.
Here, we assume that $L_i$, $N_i$ and $H_u$ 
sit at three different fixed points on the 
2D $Z_3$ orbifold.
In this case, the neutrino Yukawa couplings universally 
have an exponential suppression factor like 
eqs.(\ref{Yukawa-123}), (\ref{M-matrix}) and (\ref{Yukawa-222}).
Then, 
we can derive $(Y_\nu)_{33} =O(10^{-12})$, 
when we take the radius of the third torus as $R_3^2 = O(1000)$.
That implies that the compactification scale is smaller by 
$O(10^{-1} - 10^{-2})$ than the string scale.
(See for phenomenological aspects of such a scenario 
e.g. Refs.~\cite{Kobayashi:2004ud,Forste:2004ie,
Kobayashi:2004ya,Buchmuller:2004hv}.)

\begin{table}
  \centering
  \begin{tabular}{ccccc}
    \hline
    Class & 
    $\Delta m^2_{31}/\Delta m^2_{21}$ & 
    $\sin^2 \theta_{12}$ & $\sin^2 \theta_{23}$ & $\sin^2 \theta_{13} $ \\
    \hline
    Assignment 1 & 
    $\sim 100$ & $\lesssim 10^{-5}$ & $\lesssim 10^{-2}$ & $\lesssim 10^{-7}$\\
    Assignment 2 & 
    $\sim 100$ & $\lesssim 10^{-5}$ & $\lesssim 10^{-2}$ & $\lesssim 10^{-7}$\\
    Assignment 3 & 
    $\gtrsim 1.4$ &  &  &  \\
    Assignment 4 & 
    14 & 0.38 & 0.70 & $6.3 \times 10^{-6}$ \\
    Assignment 5 & 
    $\sim 28$ & $\le 0.09$ & & $\lesssim 10^{-2}$ \\
    Experimental values & 
    27 & 0.30 & 0.50 & 0.000 \\
    \hline
  \end{tabular}
  \caption{Characteristics of each assignment 1--5 in the Dirac
    neutrino case.
    Typical behavior of each value is described
    for combinations resulting in relatively good fits
    in a given assignment except Assignment 4.
    The row corresponding to Assignment 4 shows the best fit.
    We omit $m_e/m_\tau$ and $m_\mu/m_\tau$ because
    they can be fit in all the assignments.}
  \label{tab:char-D}
\end{table}

\subsection{Seesaw scenario}

In this subsection, considering the seesaw scenario 
of the neutrino mass matrix, 
we systematically carry out the same 
analysis as in the previous subsection.
The relevant terms in the superpotential are 
\begin{equation}
  W \supset H_u L_i (Y_\nu)_{ij} N_j - H_d L_i (Y_e)_{ij} e^c_j
  - \frac{1}{2} N_i (M_N)_{ij} N_j ,
\end{equation}
where the third term is the right-handed majorana 
neutrino mass term.
After replacing the Higgs fields with their VEVs, we obtain the mass terms,
\begin{equation}
  W \supset - \nu_i (m_D)_{ij} N_j - e_i (m_e)_{ij} e^c_j
  - \frac{1}{2} N_i (M_N)_{ij} N_j .
\end{equation}
We assume $M_N \gg m_D$, 
and in this case the heavy fields $N_i$ can be integrated out.
Then the lighter degrees of freedom have the effective mass terms,
\begin{equation}
  W \supset - \frac{1}{2} \hat{\nu}_i (\widehat{m}_\nu)_{ij} \hat{\nu}_j ,
\end{equation}
where 
\begin{equation}
  \widehat{m}_\nu = - m_D M_N^{-1} m_D^T .
\end{equation}

In string models, the natural order of mass is 
the string scale.
On the other hand, the mass scale of $M_N$ is phenomenologically 
required to be an 
intermediate scale between the string scale and 
the weak scale.
Such an intermediate scale may be obtained by VEVs of 
some fields, as a comment will be given at the end of this section.
However, such a scenario is quite model-dependent.
Thus, we assume that $M_N$ is proportional to the identity 
for simplicity.
Under such an assumption, we obtain 
\begin{eqnarray}
  \widehat{m}_\nu&=& M_N^{-1} m_D m_D^T
                  = (U_\nu) \widehat{m}^\mathrm{diag}_\nu (U_\nu)^T , \\
  \widehat{m}^\mathrm{diag}_\nu &=& M_N^{-1} ( m^\mathrm{diag}_D )^2 ,
\end{eqnarray}
where $U_\nu$ is the same diagonalizing matrix as the
one in the Dirac neutrino mass scenario.
Hence, the mixing angles are almost the same as those 
in the Dirac neutrino mass scenario.
On the other hand, neutrino mass hierarchy is enhanced 
compared with that in the Dirac neutrino mass scenario.
The results are as follows.

\vskip .3cm
{\bf Assignment 1}

Like the Dirac neutrino mass scenario, 
we can fit only the charged lepton mass ratios properly.
The neutrino mass squared difference ratio is too large, 
and mixing angles are too small.
Those results are shown in Table~\ref{tab:char-S} .

\vskip .3cm
{\bf Assignment 2}

We can fit only the charged lepton mass ratios.
The neutrino mass squared difference ratio is too large, 
and mixing angles are too small, as shown in Table~\ref{tab:char-S}.

\vskip .3cm
{\bf Assignment 3}


We can fit the charged lepton mass ratios.
However, the neutrino mass squared difference ratio
turns out to be rather large $\gtrsim 50$ except for the case that
gives $\Delta m^2_{31}/\Delta m^2_{21} = 2$.
Also, the mixing angles
are typically either too big or too small.

\vskip .3cm
{\bf Assignment 4}

Like the Dirac neutrino mass scenario,
we can properly fit both
the charged lepton mass ratios and
the neutrino oscillation data.
For example we take $(R^2_1,R^2_2) = (23,26)$ 
in the following assignment,
\begin{eqnarray}
(L_1,L_2,L_3) = (\hat T^{(1)}_2,\hat T^{(2)}_2,\hat T^{(4,1)}_2), & & 
(N_1,N_2,N_3) = (\hat T^{(2)}_2,\hat T^{(3)}_2,\hat T^{(4,1)}_2), 
\nonumber \\
(e^c_1,e^c_2,e^c_3) = (\hat T^{(1)}_3,\hat T^{(2)}_3,\hat T^{(4,1)}_3), & & 
(H_u,H_d) = (\hat T^{(2)}_2,\hat T^{}_1).
\end{eqnarray}
Then, we can realize the experimental values of 
$m_e/m_\tau$ and  $m_\mu/m_\tau$.
Furthermore, in this assignment we obtain 
\begin{eqnarray}
\frac{\Delta m^2_{31}}{\Delta m^2_{21}} = 29, & &
    \sin^2 \theta_{12} = 0.32, \nonumber \\
\sin^2 \theta_{23} = 0.48, & &\sin^2 \theta_{13} = 3.6 
\times 10^{-6}. 
\end{eqnarray}
Thus, this assignment can realize orders of 
six observables only by two parameters.
This class of assignments include several other cases leading to 
almost the same results.

\vskip .3cm
{\bf Assignment 5}


We can fit the charged lepton mass ratios and
$\Delta m^2_{31}/\Delta m^2_{21}$.
However, $\theta_{12}$ is too small
and $\theta_{23}$ is either too big or too small,
as shown in Table~\ref{tab:char-S}.

\begin{table}
  \centering
  \begin{tabular}{ccccc}
    \hline
    Class & 
    $\Delta m^2_{31}/\Delta m^2_{21}$ & 
    $\sin^2 \theta_{12}$ & $\sin^2 \theta_{23}$ & $\sin^2 \theta_{13} $ \\
    \hline
    Assignment 1  & 
    $\sim 6000$& $\lesssim 10^{-5}$ & $\lesssim 10^{-2}$ & $\lesssim 10^{-7}$\\
     Assignment 2  & 
     $\sim 7000$&$\lesssim 10^{-5}$ & $\lesssim 10^{-2}$ & $\lesssim 10^{-7}$\\
    Assignment 3  & 
    $\gtrsim 2$ &  &  &  \\
    Assignment 4 & 
    29 & 0.32 & 0.48 & $3.6 \times 10^{-6}$ \\
    Assignment 5 & 
    $\sim 28$ & $\le 0.09$ & & $\lesssim 10^{-2}$ \\
    Experimental values & 
    27 & 0.30 & 0.50 & 0.000
    \\
    \hline
  \end{tabular}
  \caption{Characteristics of each assignment 1--5 in the seesaw case.
    Typical behavior of each parameter is described
    for combinations resulting in relatively good fits
    in a given assignment.
    The row corresponding to Assignment 4 shows the best fit.
    We omit $m_e/m_\tau$ and $m_\mu/m_\tau$ because
    they can be fit in all the assignments.}
  \label{tab:char-S}
\end{table}

\vskip .3cm

As a result, we can realize the charged lepton mass ratios, 
the ratio of neutrino mass squared differences and 
the lepton mixing angles in certain cases of Assignment 4, 
when we assume that the right-handed majorana 
neutrino mass matrix $M_N$ is proportional to the 
identity matrix.
As was in the Dirac scenario,
the neutrino mass spectrum has normal hierarchy
with vanishing lightest neutrino mass.
The neutrino Yukawa coupling $(Y_\nu)_{33}$ is $O(1)$.
We need $M_N = O(10^{15})$ GeV for the seesaw mechanism
to produce the correct neutrino mass scale.

In the above analysis, we have assumed 
$(M_N)_{ij} = M_N \delta_{ij}$ for simplicity.
It is nontrivial to realize such a right-handed neutrino 
majorana mass matrix.
Majorana mass terms can be generated by 
the $(n+2)$-point couplings 
\begin{equation}
Y_{ij} N_i N_j (\phi_1 \cdots \phi_n),
\end{equation}
after $\phi_i$ develop their VEVs.\footnote{
Such VEVs may be given e.g. through anomalous U(1) breaking  
\cite{Font:1988tp}.}
Thus, the right-handed neutrino mass matrix 
depends on VEVs of scalar fields and the selection rule for 
higher dimensional operators  \cite{Kobayashi:1995py}, 
that is, quite a model-dependent feature.
In certain cases, we may obtain $(M_N)_{ij} = M_N \delta_{ij}$.
However, we, in general, obtain nontrivial majorana 
matrix $(M_N)_{ij}$ if it is derived from 
$Y_{ij} N_i N_j (\phi_1 \cdots \phi_n)$ through VEVs of 
$(\phi_1 \cdots \phi_n)$.
In such a case, assignments other than assignment 4 might
lead to realistic results.
Thus, it would be interesting to study such case somehow 
systematically.
However, that is beyond our scope.
We leave it for future study.

\section{Conclusion}

We have systematically studied the possibility for 
realizing lepton masses and mixing angles by use of 
only renormalizable couplings derived from 
$Z_6$-I heterotic orbifold models.
We have assumed one pair of up and down type 
Higgs fields.
We have found  Assignment 4 has such a possibility in 
both the Dirac neutrino mass scenario and the simple 
seesaw scenario where the right-handed majorana mass 
matrix is proportional to identity.
The resulting neutrino mass spectrum shows normal hierarchy.

It is quite non-trivial to fit six observables only by 
two parameters $R_1$ and $R_2$.
However, how to stabilize these moduli at proper values 
is an important issue to study further.
Moreover, if F-components of moduli fields contribute 
to SUSY breaking, models corresponding to Assignment 4 
may show a certain pattern of soft SUSY breaking 
parameters.
It is interesting to study their effects on 
flavor violation.
(See e.g. \cite{Chankowski:2005jh} and references therein.)

In addition to the real parts $R_i$ of the moduli $T_i$, 
mass matrices also depend on imaginary parts of $T_i$, 
although we  have fixed them to vanish in our analysis.
When such imaginary parameters are included, 
other types of assignments may lead to realistic results.
Thus, it is interesting to carry out  the same analysis 
including imaginary parts of $T_i$.

We have systematically studied all possible assignments of 
leptons and Higgs fields to twisted states, but 
not constructed an explicit heterotic orbifold model by 
fixing gauge shifts and Wilson lines.
It is also important to construct explicitly 
heterotic models corresponding to Assignment 4.

In addition, it is interesting to extend our 
analysis to other $Z_N$ and $Z_N \times Z_M$ orbifold models 
including the quark sector.
In principle, systematical studies as in this paper 
are possible for other orbifold models.
Such a study will be done elsewhere.

Although we have discussed the possibility for 
obtaining realistic Yukawa matrices from only 
stringy renormalizable couplings in this paper, 
another possibility for realistic Yukawa matrices is 
that higher dimensional operators play roles
to generate effective Yukawa couplings after symmetry 
breaking.(See e.g. \cite{nonr}.)
Particle mixing through symmetry breaking is also one 
possibility \cite{Abel:2002ih}.
It would also be interesting to study such a possibility 
somehow systematically.
In the former case, it is quite important to study 
symmetries to control higher dimensional operators in 
string models.

\section*{Acknowledgment}

JP thanks Eung Jin Chun for
discussion on neutrino oscillation data.
T.~K.\/ is supported in part by the Grant-in-Aid for
Scientific Research  (\#16028211) and
the Grant-in-Aid for
the 21st Century COE ``The Center for Diversity and
Universality in Physics'' from the Ministry of Education, Culture,
Sports, Science and Technology of Japan.
PK is supported in part by
KOSEF Sundo Grant R02-2003-000-10085-0
and KOSEF SRC program through CHEP at Kyungpook National University.

\end{document}